\documentclass{article}
\usepackage{spconf,amsmath,graphicx}
\usepackage{hyperref,amssymb,ctable,makecell,multirow,multicol,subcaption}

\title{Latency-Controlled Neural Architecture Search for Streaming Speech Recognition}
\name{Liqiang He$^1$, Shulin Feng$^1$, Dan Su$^1$, Dong Yu$^2$}
\address{  $^1$Tencent AI Lab, Shenzhen, China\\
  $^2$Tencent AI Lab, Bellevue WA, USA}
\begin{document}
\ninept
\maketitle
\begin{abstract}
Neural architecture search (NAS) has attracted much attention and has been explored for automatic speech recognition (ASR). In this work, we focus on streaming ASR scenarios and propose the latency-controlled NAS for acoustic modeling. First, based on the vanilla neural architecture, normal cells are altered to causal cells to control the total latency of the architecture. Second, a revised operation space with a smaller receptive field is proposed to generate the final architecture with low latency. Extensive experiments show that: 1) Based on the proposed neural architecture, the neural networks with a medium latency of 550ms (millisecond) and a low latency of 190ms can be learned in the vanilla and revised operation space respectively. 2) For the low latency setting, the evaluation network can achieve more than 19\% (average on the four test sets) relative improvements compared with the hybrid CLDNN baseline, on a 10k-hour large-scale dataset.
\end{abstract}
\section{Introduction}
\label{sec:intro}

Performance improvement of automatic speech recognition (ASR) systems owes much to the dedicated hand-designed model architectures. However, designing state-of-the-art neural network architectures requires a lot of expert knowledge and takes ample time. Recently, there has been rapid progress in neural architecture search (NAS) research,  which brought significant performance improvement in the computer vision field. There has been also many research work applying NAS into audio and speech processing tasks, such as acoustic scene classification \cite{li2019neural}, ASR \cite{hu2020neural, he2020learned, chen2020darts}, keyword spotting \cite{zhang2020autokws}, speaker recognition \cite{ding2020autospeech} and text to speech (TTS) \cite{luo2021lightspeech}. Among these, our prior work \cite{he2020learned} explored NAS in the LVCSR task. We adopted the differentiable framework and revised the search space for the final architecture with low complexity. The experimental results demonstrate supreme performance on a 10k-hour dataset.

In many real scenarios, low latency streaming ASR (a.k.a., online ASR) is of critical importance. Since NAS has shown great potential in our ASR experiments, a problem naturally arising is how to perform latency-controlled NAS for streaming ASR. In speech processing tasks, the latency generally depends on two aspects: the computation overhead introduced by the model complexity, and the future information needed when processing current input. There have been many works considering model complexity in NAS to achieve better speed/accuracy trade-off \cite{xu2020latency, li2020lc, berman2020aows, liu2020block}. Most of the works focus on reducing the computational overhead although limiting the lookahead frames is also essential to reduce the overall latency.

\begin{figure}[t!]
  \centering
  \includegraphics[width=0.70\linewidth]{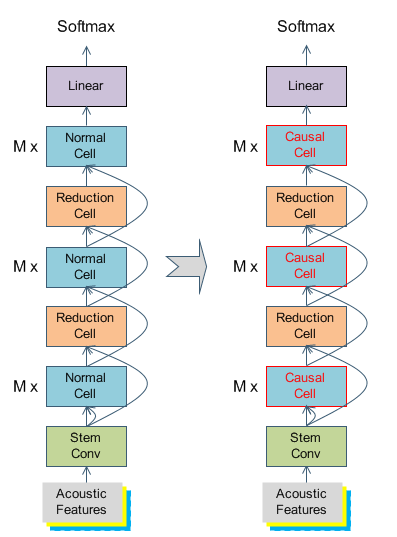}
  \vspace{-0.1cm}
  \caption{The vanilla neural architecture \cite{he2020learned} (left) without the context constraint, and the latency-controlled neural architecture (right) as proposed by our work.}
  \label{fig:arch}
  \vspace{-0.3cm}
\end{figure}

So far, there are two kinds of streaming ASR systems in industrial applications: the hybrid system and the end-to-end system. While end-to-end ASR systems have been proven to be competitive with the conventional hybrid systems, they often require to see more future contextual information, with a typical latency over 500ms, to achieve similar performance. In this work, we introduce latency constraints to the search space of NAS. We compared our latency-controlled NAS with a strong hybrid CLDNN-based system on a 10k-hour large-scale dataset. Experimental results show that the resulted evaluation network with a latency of 190ms reduced character error rate (CER) by more than 19\% (averaged on the four test sets) relatively over the hybrid CLDNN baseline with a latency of 180ms. An additional 11\% relative CER reduction can be achieved if we relax the latency to 550ms.

The contributions of this paper are as follows: 1) We propose a novel neural architecture with the normal cells replaced with the causal cells that do not depend on future context. 2) We present a revised operation space with a smaller receptive field with which low latency can be enforced in the final architecture. 3) We show that the searched architectures with the medium and low latency settings achieve significant performance improvements on the large-scale dataset, compared with the conventional hand-designed model.

\section{Vanilla Neural Architecture Search}
\label{sec:vanilla}

This section gives an overview of the NAS method in \cite{he2020learned}, which is the vanilla NAS without the latency constraint, as shown in the left half of Figure \ref{fig:arch}. The vanilla NAS follows the search space and the search strategy in DARTS \cite{liu2018darts}, except that the search process is divided into multiple stages as proposed in Progressive DARTS \cite{chen2019progressive}.

\begin{figure}[t!]
  \centering
  \includegraphics[width=0.85\linewidth]{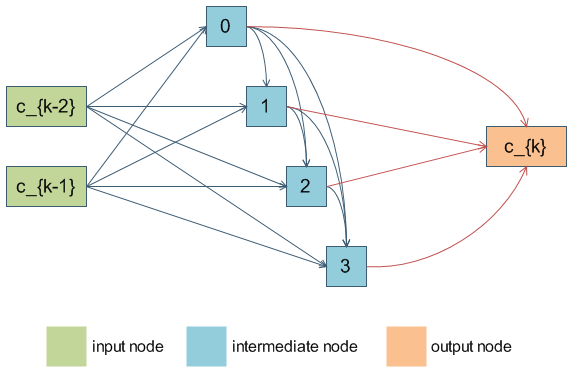}
  \vspace{-0.1cm}
  \caption{Illustration of the micro architecture of the cell in the search space. The nodes marked as $c\_\{k-2\}$ and $c\_\{k-1\}$ are the input nodes, and the node marked as $c\_\{k\}$ is the output node. The remaining ones are the intermediate nodes. Edges linked to immediate nodes are associated with the candidate operations, and the output node is defined as a depthwise concatenation of all immediate nodes.}
  \label{fig:micro_arch}
  \vspace{-0.3cm}
\end{figure}

The vanilla NAS searches for the convolutional cells as the building blocks and then stacks the searched cells to form the final architecture. Following DARTS, the vanilla NAS adopts the bi-chain-styled macro architecture \cite{xie2020weight}, i.e., each cell connects to the previous two cells, as depicted in Figure \ref{fig:arch}. The macro architecture is manually predetermined, which is composed of convolutional cells repeated many times. Two types of convolutional cells are defined to build scalable architectures \cite{zoph2018learning}. When taking in a feature map, the \textsl{normal cell} returns a feature map of the same dimension, and the \textsl{reduction cell} returns a feature map that both the height and the width are reduced by a factor of two. Two reduction cells are located at the 1/3 and 2/3 depth of the macro architecture separately. In the reduction cell, the initial operations applied to the cell’s inputs have \textbf{a stride of two} to reduce the height and width. The normal cells and the reduction cells have the same architecture respectively, but different weights for each cell.

The micro architecture \cite{xie2020weight} of each cell is a directed acyclic graph consisting of an ordered sequence of $N$ (e.g. 7) nodes, as shown in Figure \ref{fig:micro_arch}. Each node is a latent representation (e.g. a feature map in convolutional cell), and each directed edge linked to immediate nodes is associated with the candidate operations. The set of the candidate operations in DARTS is as following: \textsl{3$\times$3 and 5$\times$5 separable convolutions, 3$\times$3 and 5$\times$5 dilated separable convolutions, 3$\times$3 max pooling, 3$\times$3 average pooling, identity, and zero}. The two input nodes are marked as $c\_\{k-2\}$ and $c\_\{k-1\}$, which receive the outputs from the previous two cells, using extra convolutions to ensure the shapes matched in convolutional cell. Each intermediate node is connected to all nodes with smaller indices and input nodes, and it is computed by summing the output from all of its predecessors. The output node marked as $c\_\{k\}$ is obtained by applying a depthwise concatenation of all intermediate nodes.

DARTS proposes a differentiable network architecture search method based on bilevel optimization, which achieves remarkable efficiency improvement by several orders of magnitude, compared with conventional methods applying evolution or reinforcement learning over a discrete search space. To make the search space continuous, the categorical choice of some operation is relaxed to learning a set of continuous variables $\alpha=\{\alpha^{(i,j)}\}$, normalized with the \textsl{Softmax} function.

\begin{align}
\label{eq:relax}
\overline{o}^{(i,j)}(x)=\sum\limits_{o\in\mathcal{O}}\frac{exp(\alpha_o^{(i,j)})}{\sum_{o^{\prime}\in\mathcal{O}}exp(\alpha_{o^{\prime}}^{(i,j)})}o(x)
\end{align}
where $\mathcal{O}$ is the set of candidate operations, and each operation represents some function $o(x)$ to be applied to $x(i)$ that is the latent representation of node $i$. The architecture weight is encoded as ($\alpha_{normal}$, $\alpha_{reduce}$), where $\alpha_{normal}$ is shared by all the normal cells and $\alpha_{reduce}$ is shared by all the reduction cells.

After relaxation, a bilevel optimization algorithm is proposed to jointly optimize the architecture weight $\alpha$ as the upper-level variable and the network weights $\mathcal{\omega}$ as the lower-level variable:
\begin{align}
\label{eq:bilevel}
&\mathop{min}\limits_{\alpha} \mathcal{L}_{val}(\mathcal{\omega}^{*}(\alpha), \alpha) \\
&s.t.\quad\mathcal{\omega}^{*}(\alpha)=\mathop{argmin}_{\mathcal{\omega}}\mathcal{L}_{train}(\omega, \alpha) 
\end{align}
where $\mathcal{L}_{train}$ and $\mathcal{L}_{val}$ denote the training and the validation loss, respectively. Both losses are determined not only by the architecture $\alpha$, but also the network weights $\mathcal{\omega}$.

At the end of the search procedure, a so-called discretization process is applied to generate the final architecture. The micro architecture retains the top-$k$ (e.g. 2) strongest operations (from distinct nodes) among all non-zero candidate operations collected from all the previous nodes. The strength of an operation is defined as the continuous variables $\alpha=\{\alpha^{(i,j)}\}$ normalized with the \textsl{Softmax} function. The evaluation networks are constructed with the searched architecture. Following the scalable method \cite{zoph2018learning}, by changing the number of the normal cells, the evaluation networks can be built with the corresponding model parameters for different scenarios while keeping the number of the reduction cells unchanged.

\begin{figure*}[t!]
  \centering
  \includegraphics[width=1.0\linewidth]{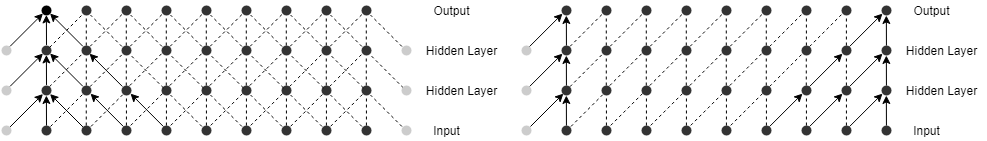}
  \vspace{-0.1cm}
  \caption{Visualization of a stack of operation layers: the vanilla operations (left) in the normal cell, and the causal operations (right) in the causal cell as proposed by our work. Gray dots are represented as the padding nodes.}
  \label{fig:operation}
  \vspace{-0.3cm}
\end{figure*}

Limited by the budget of GPU memory, DARTS has to search the architecture in a shallow (e.g. 8 cells) super-network, while the evaluation network is usually constructed with more learned cells. This discrepancy is the so-called \textsl{depth gap} between the depth of the super-network and the evaluation network, which has been proven to cause performance degradation. Progressive DARTS \cite{chen2019progressive} proposes the \textsl{search space approximation} method to alleviate the problem of the depth gap by dividing the search process into multiple stages, e.g. an initial stage, an intermediate stage, and a final stage. With each stage forwards, the super-network becomes deeper and deeper, while at the same time the number of candidate operations in the search space becomes smaller and smaller. The candidate operations with the weaker strength at the previous stage are pruned during the search process. This method makes a deeper super-network in the final stage possible with the limited computation and memory budget.

Based on the search space in DARTS, the vanilla NAS \cite{he2020learned} makes two modifications of the macro architecture for ASR tasks. At the beginning of the architecture, acoustic features and first- and second-order differences are separately assigned to the independent channels. At the ending of the architecture, the \textsl{average pooling} operation is replaced with several fully connected layers, followed by the \textsl{Softmax} layer to compute the acoustic posteriors. Two reduction cells are also adopted for ASR scenarios, as the lower frame rate technique has shown its benefit. As for the micro architecture, a revised set of the candidate operations proposed by the vanilla NAS facilitates the search algorithm to explore the convolutional architectures with low complexity for ASR tasks. Furthermore, the search process is divided into three stages as proposed in Progressive DARTS to bridge the depth gap between the search and evaluation phases.

\section{Latency-Controlled Algorithm}
\label{sec:algo}

The latency-controlled neural architecture proposed by this work modifies the neural architecture \cite{he2020learned} for streaming ASR, as illustrated in the right half of Figure \ref{fig:arch}.

Section \ref{sec:arch_intro} introduces the proposed neural architecture with the causal cells. The causal operations in the causal cell are described in section \ref{sec:operations}. The search spaces with the medium and low latency settings are presented in section \ref{sec:space}. Finally, section \ref{sec:comp} describes the latency computation of the searched architecture.

\subsection{Architecture Description}
\label{sec:arch_intro}

The macro architecture of the vanilla NAS \cite{he2020learned} is shown in the left half of Figure \ref{fig:arch}. The following issues hinder this architecture to be applied in the streaming scenarios: 1) Both the normal and reduction cells are designed without the context constraint, and the receptive field grows with the number of the stacked cells. 2) Concerning the scalable method \cite{zoph2018learning}, the latency of the evaluation network changes with the number of the stacked cells required by ASR applications. The evaluation networks are generated with very high and uncertain latency, so it's almost impossible to be applied in streaming ASR applications.

To solve the latency problems of the vanilla neural architecture, we propose a novel macro architecture with the normal cells replaced with the causal cells, as shown in the right half of Figure \ref{fig:arch}. Different from the normal cell, operations in the causal cell only depend on the left (history) context. After this modification, the latency of the proposed neural architecture only comes from the reduction cells. Once the structure of the reduction cell is learned, the latency of the evaluation networks is determined. Thus, we only need to control the latency of the reduction cells in the search phase for online streaming applications.

\subsection{Causal Operations}
\label{sec:operations}

The vanilla candidate operations \cite{he2020learned} with the right (future) context dependency in the convolutional cell are as follows: \textsl{[average pooling, max pooling, separable convolution, dilated separable convolution]}. The receptive field, including the right context, grows linearly with the number of the stacked operation layers, as depicted in the left half of Figure \ref{fig:operation}.

Concerning the context constraint, the causal convolution is proposed by WaveNet \cite{oord2016wavenet}, and a masked convolution, equivalent to the causal convolution, is realized in \cite{van2016pixel}. Considering all the candidate operations, we adopt the causal convolution \cite{oord2016wavenet} for the following operations: \textsl{separable convolution}, and \textsl{dilated separable convolution}. Furthermore, we extend this causal method to the remaining operations: \textsl{average pooling}, and \textsl{max pooling}. Through this, we make sure that all the modified operations, namely causal operations, in the search space do not depend on any of the right contexts, as shown in the right half of Figure \ref{fig:operation}. The causal cells composed of the causal operations thus depend on none of the right contexts.

\begin{figure*}[tp!]
    \centering
    \vspace{-0.1cm}
    \begin{subfigure}{0.48\textwidth}
         \centering
         \includegraphics[width=0.96\textwidth]{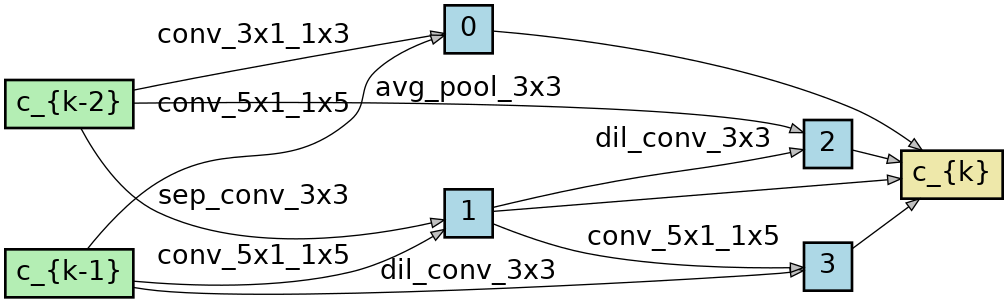}
         \caption{Causal cell learned in the low latency search space.}
         \label{subfigure-1}
    \end{subfigure}
    \hfill
    \begin{subfigure}{0.48\textwidth}
         \centering
         \includegraphics[width=0.96\textwidth]{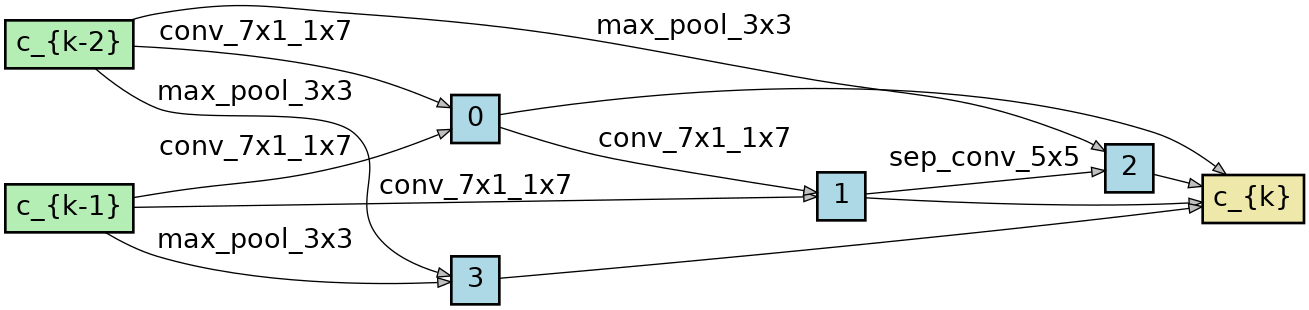}
         \caption{Causal cell learned in the medium latency search space.}
         \label{subfigure-2}
    \end{subfigure}
    \vfill
    \begin{subfigure}{0.48\textwidth}
         \centering
         \includegraphics[width=0.96\textwidth]{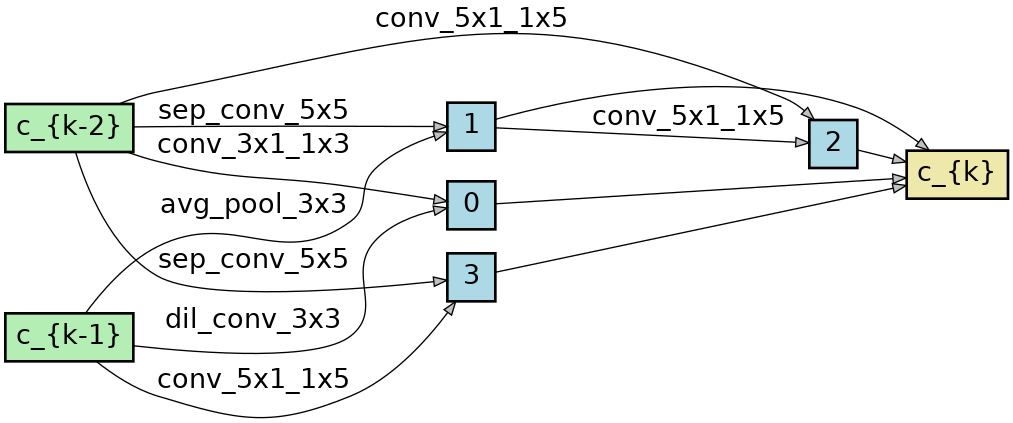}
         \caption{Reduction cell learned in the low latency search space.}
         \label{subfigure-3}
    \end{subfigure}
    \hfill
    \begin{subfigure}{0.48\textwidth}
         \centering
         \includegraphics[width=0.96\textwidth]{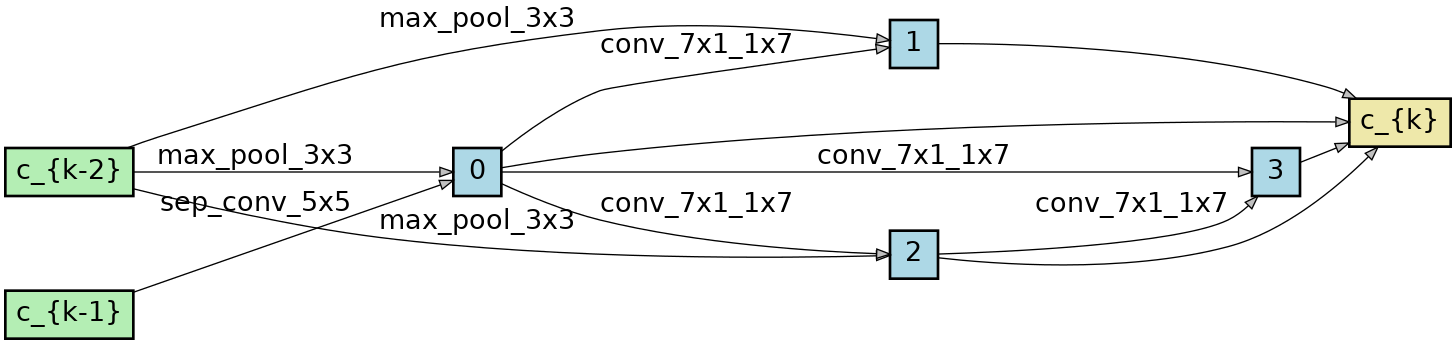}
         \caption{Reduction cell learned in the medium latency search space.}
         \label{subfigure-4}
    \end{subfigure}
\vspace{-0.1cm}
\caption{(a) and (c) are the cells (denoted as ASRNET-C with a latency of 190ms) learned in the low latency search space. (b) and (d) are the cells (denoted as ASRNET-D with a latency of 550ms) learned in the medium latency search space.}
\label{fig:cell}
\vspace{-0.3cm}
\end{figure*}

\subsection{Medium and Low Latency Search Spaces}
\label{sec:space}

The latency of the reduction cell is determined by the following aspects: 1) The micro architecture of the searched cell is a principal factor to influence the total latency. 2) The right context dependency of each finally selected operation is another important consideration. Theoretically, the total latency is computed by the summation of the latency of each selected operation in the deepest path of the searched reduction cell.

Concerning the first issue, a range of micro architectures with different latencies can be searched from the same search space with random initial seeds. Theoretically, we can run the architecture search many times and choose the micro architecture with low latency. However, we found empirically that micro architectures with good performance always have high latency.

As for the second issue, the search algorithm automatically searches for the optimal architectures with the corresponding latency based on the different operation spaces. The architecture with low latency can be learned in the dedicated operation space. For comparison, the operation space in the vanilla NAS is adopted for the medium latency search space, which is shared in both the causal and reduction cells. All the candidate operations in the causal cell are with the context constraint as stated in section \ref{sec:operations}. The optimal architecture for the medium latency setting is searched in the cited search space, as shown in the right half of Figure \ref{fig:cell}.

For the low latency setting, the revised operation space are listed as following: \textbf{\emph{[zero, 3$\times$3 max pooling, 3$\times$3 average pooling, 3$\times$3 separable convolution, 5$\times$5 separable convolution, 3$\times$3 dilated separable convolution, 3$\times$1 then 1$\times$3 convolution, 5$\times$1 then 1$\times$5 convolution]}}. Compared with the search space in the vanilla NAS, the operations \textsl{[5$\times$5 dilated separable convolution, 7$\times$1 then 1$\times$7 convolution]} are replaced with the operations \textsl{[3$\times$1 then 1$\times$3 convolution, 5$\times$1 then 1$\times$5 convolution]} for the smaller receptive fields. The convolutions in the candidate operations follow an ordering \cite{liu2018darts, zoph2018learning} of \textsl{ReLU}, \textsl{convolution} operation and \textsl{Batch Normalization}. Sequential modules with a \textsl{ReLU-Conv-BN} order are applied only once to the \textsl{separable convolution} in the revised operation space, while they are stacked twice in the vanilla NAS, leading to a double receptive field. As shown in the left half of Figure \ref{fig:cell}, the optimal architecture for the low latency setting is searched in the revised search space.

\subsection{Latency Computation}
\label{sec:comp}

It should be noted that we use \textsl{algorithmic latency} induced by the right context dependency. There are many factors to influence the latency of streaming ASR systems. For example, the computation delay of the neural network, and the delay of label emission in the CTC (connectionist temporal classification) criterion. The former is possibly handled by the model compression method, and the latter could be possibly avoided by adopting the CE (cross-entropy) criterion. However, these are beyond what we discuss in this paper.

Some considerations in the latency computations are listed as following: 1) The \textsl{convolution} with the kernel \textsl{3$\times$3} in the \textsl{stem convolution} layer produces extra 10ms latency. 2) In the reduction cell, the initial operations applied to the input nodes have a stride of two to reduce the height and width of the feature maps. 3) Based on issue 2), the reduction cells share the searched architecture, so the latency of the second reduction cell is twice that of the first one. 4) The sequential modules with a \textsl{ReLU-Conv-BN} order are applied to the \textsl{separable convolution} twice in the medium latency search space, while only once to the \textsl{separable convolution} in the low latency search space. As shown in the lower left of Figure \ref{fig:cell}, the path with the biggest receptive field contains the operations in sequence: \textsl{5$\times$5 separable convolution}, \textsl{5$\times$1 then 1$\times$5 convolution}. The first \textsl{separable convolution} produces 20ms latency, while the second \textsl{convolution} introduces 2 $*$ 20ms for a stride of two in the first operation. The latency of the first reduction cell is summed to 60ms, and the latency of the second reduction cell could be computed to 120ms, given issue 3). Plus the extra 10ms, the total 190ms latency is induced for the low latency setting. As shown in the lower right of Figure \ref{fig:cell}, the path with the biggest receptive field includes: \textsl{5$\times$5 separable convolution}, \textsl{7$\times$1 then 1$\times$7 convolution}, \textsl{7$\times$1 then 1$\times$7 convolution}. Considering issues 2) and 4), the first \textsl{separable convolution} produces 20ms $+$ 2 $*$ 20ms latency, where the 20ms latency is introduced by the first sequential modules, and the 2 $*$ 20ms latency is produced by the second sequential modules. So, the \textsl{separable convolution} totally generates 60ms latency. The second \textsl{7$\times$1 then 1$\times$7 convolution} introduces 2 $*$ 30ms latency for a stride of two in the first operation, so does the third \textsl{convolution}. The latency of the first reduction cell is summed to 180ms, and the latency of the second reduction cell could be computed to 360ms, given issue 3). Plus the extra 10ms, the total 550ms latency is induced for the medium latency setting.

\section{Experiments}
\label{sec:exper}

\subsection{Datasets}

AISHELL-1 \cite{bu2017aishell} is used as the small proxy dataset in the search phase. To verify the transferability of the searched architecture, two larger corpora are used in the evaluation phase: 1) AISHELL-2 \cite{du2018aishell} dataset, which contains 1000 hours of speech data from 1991 speakers, is used as the medium-scale dataset. 2) A 10k-hour dataset \cite{you2020dfsmn} is used as a large-scale dataset, collected from more than \textbf{ten industrial scenarios}, e.g. conversation speech, children's speech, Chinese dialect data, spontaneous speech, voice search data, and so on. Further, AISEHLL-1 and AISEHLL-2 are augmented with 2-fold speed perturbation \cite{ko2015audio} in the experiments. To make the performance evaluation more comprehensive, results are reported on three types of test sets, which consist of hand-transcribed anonymized utterances extracted from reading speech (1001 utterances), conversation speech (1538 utterances), and spontaneous speech (2952 utterances). The above test sets are referred to as Read, Chat, and Spon respectively for short. Additionally, the AISHELL-2 development set (2500 utterances, short for DEV), recorded by high fidelity microphone, is used as the test set to provide a public benchmark.

\subsection{Training setup}
For acoustic features, we use 40-dimensional filterbanks extracted with a 25ms window with a stride of 10ms, extended with temporal first- and second-order differences. The length of the training utterances is filtered by a maximum limit of 1024 frames, and 4-frames-aligned padding was made for the length of each utterance to be fit for the two reduction layers. The CTC (connectionist temporal classification) learning framework is applied in the experiments, and the models are trained with multiple GPUs with BMUF \cite{chen2016scalable}. We use the CI-syllable-based model units: 1394 Mandarin syllables, 39 English phones, and a blank. A beam search algorithm, using weighted finite-state transducers (WFSTs), performs the first-pass decoding with a pruned 5-gram language model and a second-pass rescoring with RNN model. The performance results are measured by Character Error Rate (CER).

\subsection{Architecture Search}

Refer to \cite{he2020learned} for most of the hyperparameters and configurations of the search phase. We randomly split the training set of the AISHELL-1 dataset into two equal subsets, one half for learning network parameters and the other half for tuning the architecture parameters. We divide the search process into three stages \cite{chen2019progressive}. In each stage, only network parameters are updated in the first 6 epochs, and both architecture and network parameters are alternately trained in the remaining 9 epochs. Following the search space regularization \cite{chen2019progressive}, the dropout probabilities are applied on the candidate operations \textsl{[dilated separable convolution, average pooling]}. Three sets of the dropout [0.00, 0.00, 0.00], [0.00, 0.05, 0.10], [0.00, 0.10, 0.20] are used, and the values in each set correspond to three stages, respectively. Besides, the discovered causal cells are restricted to keep at most 2 \textsl{average pooling}. To alleviate the influence of randomness, the search process is repeated 3 times with different seeds for each dropout probability setting. Two optimal architectures are searched as follows: \textsl{ASRNET-C} with a low latency of 190ms as shown in the left half of Figure \ref{fig:cell}, and \textsl{ASRNET-D} with a medium latency of 550ms as depicted in the right half of Figure \ref{fig:cell}. The search process takes around 96 hours on 8 Tesla P40 GPUs.

\begin{table}[tp!]
\centering
\caption{Token accuracies (Acc) and evaluation costs (Cost) of the networks on the AISHELL-1 dataset. Abbreviations: L is the number of cells, and C is the initial number of channels.}
\label{tab:small}
\vspace{-0.1cm}
\begin{tabular}{c|c|c|c}
\specialrule{.13em}{0em}{0em}
\makecell{Small} & \makecell{Params\\($M$)} & \makecell{Acc\\($\%$)} & \makecell{Cost\\($hours$)} \\
\hline
\makecell{ASRNET-C (L=17,C=25)} & \makecell{6.4} & \makecell{90.93} & \makecell{13.6} \\
\hline
\makecell{ASRNET-D (L=17,C=22)} & \makecell{6.5} & \makecell{91.27} & \makecell{12.4} \\
\specialrule{.13em}{0em}{0em} 
\end{tabular}
\vspace{0.1cm}
\end{table}

\begin{table}[tp!]
\centering
\caption{Comparison with medium-sized CLDNN on the AISHELL-2 dataset. \textsl{Rel Imp} refers to Relative Improvement.}
\label{tab:medium}
\vspace{-0.1cm}
\begin{tabular}{c|c|c|c}
\specialrule{.13em}{0em}{0em}
\makecell{Medium} & \makecell{Params\\($M$)} & \makecell{CER\\($\%$)} & \makecell{\textsl{Rel Imp} \\($\%$)}\\
\hline
\makecell{CLDNN (90ms)} & \makecell{37.0} & \makecell{8.59} &  $/$ \\
\hline
\makecell{CLDNN (180ms)} & \makecell{37.0} & \makecell{7.45} &  - \\
\hline
\makecell{ASRNET-C (L=26,C=30)} & \makecell{11.1} & \makecell{6.28} & \makecell{\textsl{15.70}}  \\
\hline
\makecell{ASRNET-D (L=26,C=30)} & \makecell{14.4} & \makecell{5.71} & \makecell{\textsl{23.36}}\\
\specialrule{.13em}{0em}{0em} 
\end{tabular}
\vspace{-0.3cm}
\end{table}

\subsection{Architecture Evaluation}

On the AISHELL-1 dataset, the small evaluation networks contain 17 cells with initial channels of 25 and 22 for \textsl{ASRNET-C} and \textsl{ASRNET-D} respectively. The reason for different configurations is to make the model parameters of the two networks similar and ensure the fairness of comparison. Both networks are trained from scratch for 20 epochs with batch size 4 and the token accuracies are measured on the test set. As shown in Table \ref{tab:small}, the performance of  \textsl{ASRNET-D} with higher latency is slightly better than  \textsl{ASRNET-C}. The training tasks are performed on 8 Tesla P40 GPUs.

The widely used hybrid model CLDNN \cite{sainath2015convolutional, amodei2016deep} is used as the baseline. In streaming ASR scenarios with low latency, CLDNN architecture is one of the most powerful models for industrial applications, while other neural architectures \cite{wu2020transformer, chen2021transformer} usually relax the latency for more future contextual information to achieve competitive performance. For the very low latency setting, CLDNN starts with two convolution layers with the kernel size \textsl{5$\times$5}, and each layer is followed by the \textsl{max pooling} layer with the kernel size \textsl{2$\times$1} and a stride of \textsl{2$\times$1}. For the low latency setting, only the kernel size of the convolution layer is changed to \textsl{11$\times$5}. The first value of the kernel is applied to the time axis and the second one to the frequency axis. Two settings are configured as follows: very low latency of 90ms, and low latency of 180ms.

On the AISHELL-2 dataset, the medium-sized evaluation networks contain 26 cells with initial channels of 30 for \textsl{ASRNET-C} and \textsl{ASRNET-D}. The networks are trained from scratch for 15 epochs with batch size 4. The medium-sized CLDNN consists of 7 LSTM layers with 1024 nodes in each layer. As seen in Table \ref{tab:medium}, \textsl{ASRNET-C} has achieved more than 15\% relative improvements for the low latency setting, and \textsl{ASRNET-D} has achieved more than 23\% relative improvements for the medium latency setting, compared with the CLDNN baseline with a latency of 180ms. As for the configurations of the networks, the initial configurations are the key points for the fairness of performance comparisons with the medium-size parameters, based on the empirical experiments. The training processes are accelerated by using 24 Tesla P40 GPUs.

On a 10k-hour large-scale dataset, the large evaluation networks contain 26 cells with initial channels of 56 and 50 for \textsl{ASRNET-C} and \textsl{ASRNET-D} respectively. Except for the model parameters, the \textsl{GPU memory budget} is taken into account for different initial configurations. The networks are trained from scratch for 7 epochs with batch size 4. The large-sized CLDNN consists of 7 LSTM layers with 1536 nodes in each layer. As shown in Table \ref{tab:large}, for the low latency setting, the network has achieved more than 19\% (average on the four test sets) relative improvements compared with the CLDNN baseline. More than 11\% relative improvements have been achieved by the network with the medium latency setting, compared with the low latency setting. Ablation study indicates that the CLDNN baseline with the low latency setting has obvious performance advantages over the very low latency setting. The training process takes around 7 days on 24 Tesla V100 GPUs.

\begin{table}[tp!]
\centering
\caption{Comparison with large-sized CLDNN on a 10k-hour dataset. CERs are measured on the four test sets.}
\label{tab:large}
\vspace{-0.1cm}
\begin{tabular}{c|c|c|c|c|c}
\specialrule{.13em}{0em}{0em}
\makecell{Large} & \makecell{Params\\($M$)} & \makecell{Read\\(\%)} & \makecell{Chat\\(\%)} & \makecell{Spon\\(\%)} & \makecell{DEV\\(\%)} \\
\hline
\makecell{CLDNN \\ (90ms)} & \makecell{55.12} & \makecell{3.35} & \makecell{32.52} & \makecell{31.75} & \makecell{6.31} \\
\hline
\makecell{CLDNN \\ (180ms)} & \makecell{55.15} & \makecell{2.95} & \makecell{28.78} & \makecell{30.87} & \makecell{5.73} \\
\hline
\makecell{ASRNET-C\\(L=26,C=56)} & \makecell{32.8} & \makecell{2.22} & \makecell{25.45} & \makecell{24.26} & \makecell{4.54} \\
\hline
\makecell{ASRNET-D\\(L=26,C=50)} & \makecell{35.8} & \makecell{1.74} & \makecell{22.52} & \makecell{22.95} & \makecell{4.27} \\
\specialrule{.13em}{0em}{0em} 
\end{tabular}
\vspace{-0.3cm}
\end{table}

\section{Conclusions}

In this paper, we focus on developing a streaming ASR model with NAS. We modify the macro architecture of NAS to restrict future context, and to search in the revised operation space to generate the architecture with low latency. We tested our method on a 10k-hour large-scale dataset. Experimental results demonstrated the great potential of our proposed approach. For the low latency setting, the searched model achieves 19\% relative improvement compared with a CLDNN hybrid baseline model. Future work includes testing our method with transducer-based end-to-end streaming ASR systems.

\bibliographystyle{IEEEbib}
\bibliography{strings,refs}

\end{document}